\documentclass{article}
\usepackage[english]{babel}
\usepackage[utf8x]{inputenc}
\usepackage{graphicx}
\usepackage{epsfig}
\usepackage{color}
\usepackage{graphicx}
\usepackage{nameref}
\usepackage{epstopdf, epsfig}
\usepackage{amsmath,amsfonts,amssymb}

\usepackage{xcolor}
\usepackage{hyperref}
\newcommand{\be}{\begin{equation}}
\newcommand{\ee}{\end{equation}}

\begin{document}
\title{Magnetic field dynamics in presence of  thermal diffusion and Hall conductivity}

\author{G.S. Bisnovatyi-Kogan, M.V. Glushikhina \\
Space Research Institute of Russian Academy of Sciences,\\ Profsoyuznaya 84/32, Moscow 117997, Russia,\\ 
Email: gkogan@iki.rssi.ru}
\date{}

\maketitle

\begin{abstract} 
 Anisotropy of kinetic coefficients in presence of a magnetic field is
 represented by Hall currents, which appear in a collisional medium due to action of the Lorentz force on the charged particles between collisions.  
 We derive equations, describing dynamics of the magnetic field in presence of {thermal diffusion} with Hall currents, using a standard electrodynamic consideration. The influence of the Hall currents, at presence of {thermal diffusion}, on the magnetic field structure  is considered in simple models. 
 {The equation is derived, which includes  additional term for a seed magnetic field creation by the mechanism known as  "Biermann battery" in the unmagnetized plasma.}       
\end{abstract}

	\section{Introduction}

		In presence of a magnetic field,  kinetic coefficients, describing charged particles, became anisotropic. It is related to  plasma, as well as to  metals. Such anisotropy  represents appearance of Hall currents, which appear in a collisional plasma,                              due to action of the Lorentz force on the charged particles between collisions. 
	The characteristic feature of Hall currents is their property of preserving entropy, and absence of a heat production. Therefore the 
	influence of Hall currents on a 
	magnetic field damping  may happen only indirectly, by changing the magnetic field structure.
	
	Transport coefficients in the non-degenerate magnetized plasma had been found in \cite{brag57, brag65}  using solution of the Boltzmann kinetic equation by Chapmen-Enskog method \cite{chcow}. {Thermal diffusion} and Hall currents had been taken into account in these papers, together with analysis of different physical effects. The equation for the excitation of a seed magnetic field in the {unmagnetized plasma}, obtained there, is connected with a pressure gradient term in the diffusion vector ${\bf d_e}$  \cite{chcow, brag65}
	
	\begin{eqnarray}
		\label{dv}
		{\bf d_e}=\frac{e}{kT}\large[{\bf E}+\frac{1}{c}[{\bf v}\times{\bf B}]\large]+\frac{{\bf\nabla}P_e}{P_e}\\ \nonumber
		=\frac{e}{kT}\large[{\bf E}+\frac{1}{c}[{\bf v}\times{\bf B}]+\frac{{\bf\nabla}P_e}{e n_e}\large]
	\end{eqnarray}
	
	
	We have derived here the equation for   
	the excitation of a seed magnetic field in presence of {thermal diffusion}, and have obtained the additional term in the equation for seed field excitation. The creation of a seed magnetic field is a very important astrophysical problem.  The presence of magnetic field everywhere in the universe is explained by different mechanisms of its amplification, but these processes could happen only after creation of initial seed field in the initially {unmagnetized} universe,   
	
	Development of a laboratory astrophysics  is  aimed  to model astrophysical events in a terrestrial laboratory.
	MHD simulations with Hall current had been performed in  \cite{shaikhislamov2013}, to simulate  the laboratory experiments.  
	Numerical calculations of the magnetic field evolution in neutron star crust had been done in \cite{ger2002, pons2007} in presence of the Hall effect. 
	Here similar equations are derived in presence of a temperature gradient, external action, and {thermal diffusion}.

	We have separated magnetic and electric fields on {self-field} and external one, depending on their origin, {self electric} currents and {electric} charge distribution, and external source. Those could be the {electric} field produced by battery, accumulators etc., and magnetic field originated by external currents, or external non-vortex magnetic field.
	
	In presence of {thermal diffusion}, the additional channel for creation of the seed magnetic field in the {unmagnetized} media is found, which is needed for the action of the mechanism known as "Biermann battery" \cite{bier1, bier2}. 		
	
	Simple models are considered for illustration of Hall current's influence on magnetic filed structure, including the action of {thermal diffusion}.

	\section{MHD equations in presence of Hall conductivity}
	
{The fluid-Maxwell equations include the Maxwell equations} without a displacement current \cite{cowling} and a generalized Ohm law for anisotropic conductivity in presence of the magnetic field in the form: 

\begin{eqnarray}
	\label{1m}
	\frac{1}{c}\frac{\partial\bf B}{\partial t}=-{\bf \nabla\,
		\times\,E}, \qquad {\bf\nabla \cdot B} =0, \\
	\label{2m}	
	{\bf j}=\frac{c}{4\pi} {\bf\nabla\,\times\,B},\\
	\label{3m}	
	{\bf j}=\sigma_{ij}{{\bf E'}}.
\end{eqnarray}
Anisotropic conductivity tensor $\sigma_{ij}$ has a relatively
simple form in the first Chapman-Enskog approximation, ${\bf E'}$ is the the {electric} field in the co-moving coordinate system.
The expression in (\ref{3m}) may be
written in the form \cite{urpin, pons2019}:

\begin{eqnarray}
	\label{1j}	
	{\bf j}=\frac{\sigma_0}{1+\omega^2 {\tau_e}^2}
	\biggl[{\bf E'}+\frac{\omega^2 {\tau_e}^2}{B^2}{\bf (B\cdot E')B} -\frac{\omega\tau_e}{B} \bf{[B\times E']}\biggr].
\end{eqnarray}
For a non-relativistic, non-degenerate, fully ionized plasma we have the following scalar coefficients of its conductivity $\sigma_0$ in the absence of the magnetic field, see \cite{brag65, bkg, glu}: 

\begin{equation}
	\label{12jt}
	\sigma_0=\frac{32 e^2 n_e}{3 \pi m_e} \tau_e,\,\, {\rm with}\,\, \tau_e = \frac{3}{4}\sqrt{\frac{m_e}{2 \pi}}\frac{(k T)^{3/2}}{Z^2 e^4 n_N \Lambda}. 
\end{equation}	
Here
$\omega=(eB/m_e c)$ is the electron Larmor frequency,  
$\tau_e$ is a time between electron-ion collision, in
Lorentz approximation;  $n_e$, $n_N$ are concentrations of the electrons and nuclei with atomic number Z,  $\Lambda$ is a Coulomb logarithm,  ${\bf E}$, ${\bf E'}$ are electric fields in the laboratory and comoving coordinate system, with account of the pressure gradient term, respectively . The comoving system is moving with velocity ${\bf v}$ relative to the laboratory system together with matter, so that \cite{cowling, brag65}

\begin{equation}
	\label{2j}
	{\bf E'}= {\bf E}+\frac{1}{c}{\bf [v\times B]}+\frac{{\bf\nabla} P_e}{e n_e}.
\end{equation}
From equations \eqref{1m}-\eqref{1j} follows the expression for ${\bf E'}$ as a function of ${\bf j,\, B}$, with account of Hall currents, in the form  \cite{cowling, lpit} 	
\begin{equation}
	\label{4e}
	{\bf E'}= \frac{\bf j}{\sigma_0}-\frac{\omega\tau_e}{B\sigma_0}
	{\bf[B\times j]} = \frac{c}{4\pi \sigma_0}{\bf \nabla \times B}-\frac{c\, \omega\tau_e }{4\pi\sigma_0 B}{\bf B\times[\nabla\times B]}.
\end{equation}  
Substituting (\ref{4e}) into (\ref{1m}), with account of (\ref{2j}), leads to the equation for the evolution of a magnetic
field 
\cite{brag65, young, urpin, rhein2002, pons2007, pons2019} at anisotropic electric conductivity due to Hall current, as

\begin{eqnarray}
	\label{4f}
	\frac{\partial{\bf B}}{\partial t}={\bf \nabla}\times 
	[{\bf v\times B}] + 
	\frac{c^2}{4\pi \sigma_0}  {\bf \nabla^2  B} \qquad \qquad \qquad \\
	+\frac{c^2\,\omega\tau_e}{4\pi\sigma_0 B} {\bf \nabla}\times[{\bf B\times[\nabla\times B]}]-\frac{ck}{e n_e}[{\bf \nabla}n_e\times {\bf \nabla} T]. \nonumber
\end{eqnarray}  

The hydrodynamic part of MHD equations \cite{lleldyn} (continuity, motion energy, and heat propagation) should be added to the Eqs.(\ref{4f}), at Ohm's law (\ref{1j}), for solving problems appearing in modeling of events in the space and in a laboratory. 

The equation (\ref{4f}) is describing the evolution of the magnetic field in the medium moving with {mass center velocity} $\bf v$ relative to the laboratory frame.

It is interesting  to compare  (\ref{4f}) with the equation (39) from the paper  \cite{gold}, where the first term to the right has the same dependence on $\bf B$, as the second  term from the end in (\ref{4f}), but has a different coefficient.
The authors of \cite{gold} have called it as "Hall drift". Actually this term is connected with Lorentz transformation of the {electric} field {\bf E} from the laboratory frame to the 
co-moving one \cite{cowling}. The Hall effect is connected with the anisotropic {electric} conductivity in the conducting medium in presence of the magnetic field, which is included in the generalized Ohms law \cite{chcow}, given here in Eq.(\ref{1j}).
In \cite{gold} the equation (39) is applied to the metallic static body with the electric current $\bf j$, which is obtained from the 
Lorentz transformation equation by identifying 
${\bf j}$= $e{\bf v}$. 
This procedure is not correct, because the center mass ${\bf v}$, related to the moving frame, is actually zero. 
According to (\ref{4f}), in the ideal plasma  with infinite $\sigma_0$, and absence of the pressure gradient term, there is no {dissipation} of the magnetic field, no influence of the Hall currents on its behavior, and in the co-moving frame the electric field is zero. In \cite{gold} the so-called "Hall drift" term remains non-zero in the ideal plasma, what leads to artificial behavior of the magnetic field in it.
	
	\section{Equations in the presence of thermal diffusion and external electro-moving force (EMF) }
	 	With account of {thermal diffusion} and {\bf external {electric} field 
	 } ${\bf E_{ex}}$, see
	 \cite{cowling, chcow, bk64, brag57, brag65}
	 the Ohm's law has the following form:
	 
	 \begin{eqnarray}
	 	\label{1jt}	
	 	{\bf j}=\frac{\sigma_0}{1+\omega^2 {\tau_e}^2}
	 	\biggl[{\bf E'}+\frac{\omega^2 {\tau_e}^2}{B^2}{\bf (B\cdot E')B}
	 	-\frac{\omega\tau_e}{B} \bf{[B\times E'}]\biggr]
	 	\nonumber \\
	 	\label{2jt}
	 	+\frac{\sigma_0}{1+\omega^2 {\tau_e}^2}
	 	\biggl[{\bf E_{ex}}+\frac{\omega^2 {\tau_e}^2}{B^2}{\bf (B\cdot E_{ex})B}
	 	-\frac{\omega\tau_e}{B} \bf{[B\times E_{ex}]}\biggr] \\	\nonumber     
	 	+\frac{\lambda_0}{1+\omega^2 {\tau_e}^2}
	 	\biggl[\bf{\nabla T}+\frac{\omega^2 {\tau_e}^2}{B^2}{\bf (B\cdot  \nabla}T){\bf B} -\frac{\omega\tau_e}{B} \bf{[B\times \nabla T}]\biggr].
	 \end{eqnarray}  
	 Here $\lambda_0$ is a scalar {thermal diffusion} coefficient, which for a non-relativistic, non-degenerate plasma in absence of a magnetic field, in Lorenz approximation is written as
	 
	 \begin{equation}
	 	\label{12jt1}
	 	\lambda_0= e n_e \mu^{(1)}=\frac{16ken_e}{\pi m_e}\tau_e.
	 \end{equation}
	 The linear equations, connecting the components of the {electric}
	 current $j_i$, {electric} field in the co-moving frame $E'_i$, external {electric} field 
	 $E_{ex,i}$,	and  temperature gradient $\frac{\partial T}{\partial x_i}$ are
	 written as ($i=x,y,z$):
	 \begin{eqnarray}
	 	\frac{1+\omega^2 {\tau_e}^2}{\sigma_0} j_x=
	 	E'_x+ E_{ex, x}  +\frac{\omega^2 {\tau_e}^2}{B^2}[B_x (E'_x+E_{ex, x}) \\ \nonumber +B_y (E'_y+E_{ex, y})
	 	\label{3jt}
	 	+B_z( E'_z+E_{ex, z})]B_x   \\ \nonumber
	 	-\frac{\omega\tau_e}{B}[B_y (E'_z+E_{ex, z})  -B_z (E'_y+E_{ex, y})] \\ \nonumber
	 	+\frac{\lambda_0}{\sigma_0}\biggl[\frac{\partial T}{\partial x}  +\frac{\omega^2 {\tau_e}^2}{B^2}(B_x\frac{\partial T}{\partial x} +B_y \frac{\partial T}{\partial y} \\ \nonumber+B_z \frac{\partial T}{\partial z})B_x  
	 	-\frac{\omega\tau_e}{B}(B_y \frac{\partial T}{\partial z}-B_z \frac{\partial T}{\partial y})\biggr], 
	 \end{eqnarray}
	 
	 \begin{eqnarray}
	 	\frac{1+\omega^2 {\tau_e}^2}{\sigma_0} j_y=
	 	E'_y+E_{ex, y}  +\frac{\omega^2 {\tau_e}^2}{B^2}[B_x (E'_x+E_{ex, x})\\ \nonumber+B_y( E'_y+E_{ex, y}) 
	 	\label{4jt} 
	 	+B_z (E'_z+E_{ex, z})]B_y \\ \nonumber	
	 	-\frac{\omega\tau_e}{B}[B_z (E'_x+E_{ex, x})-B_x (E'_z+E_{ex, z})] \\  \nonumber
	 	+\frac{\lambda_0}{\sigma_0}\biggl[\frac{\partial T}{\partial y} +\frac{\omega^2 {\tau_e}^2}{B^2}(B_x\frac{\partial T}{\partial x}+B_y \frac{\partial T}{\partial y}\\ \nonumber+B_z \frac{\partial T}{\partial z})B_y 
	 	-\frac{\omega\tau_e}{B}(B_z \frac{\partial T}{\partial x}-B_x \frac{\partial T}{\partial z})\biggr], 	
	 \end{eqnarray}

	 \begin{eqnarray}
	 	\frac{1+\omega^2 {\tau_e}^2}{\sigma_0} j_z=
	 	E'_z+E_{ex, z} + \frac{\omega^2 {\tau_e}^2}{B^2}[B_x (E'_x+E_{ex, x})\\  \nonumber +B_y (E'_y+E_{ex, y}) 
	 	\label{5jt}
	 	+B_z (E'_z+E_{ex, z})]B_z \\ \nonumber
	 	-\frac{\omega\tau_e}{B}[B_x(E'_y+E_{ex, y})-B_y (E'_x+E_{ex, x})] 	\\  \nonumber
	 	+\frac{\lambda_0}{\sigma_0}
	 	\biggl[\frac{\partial T}{\partial z} +\frac{\omega^2 {\tau_e}^2}{B^2}(B_x\frac{\partial T}{\partial x}+B_y \frac{\partial T}{\partial y} \\  \nonumber +B_z \frac{\partial T}{\partial z})B_z 
	 	-\frac{\omega\tau_e}{B}(B_x \frac{\partial T}{\partial y}-B_y \frac{\partial T}{\partial x})\biggr]. 
	 \end{eqnarray}
	 Introducing
	 
	 \begin{eqnarray}
	 	\label{6jt}
	 	\tilde E_x=E'_x+E_{ex, x}+\frac{\lambda_0}{\sigma_0}\frac{\partial T}{\partial x},\quad \tilde E_y=E'_y+E_{ex, y}+\frac{\lambda_0}{\sigma_0}\frac{\partial T}{\partial y}, \\ \nonumber \quad \tilde E_z=E'_z+E_{ex, z}+\frac{\lambda_0}{\sigma_0}\frac{\partial T}{\partial z}, \qquad \qquad
	 \end{eqnarray}
	 
	 we obtain the solution of the system (\ref{3jt})-(\ref{5jt}) in the form:
	 
	 \begin{eqnarray}
	 	\label{7jt}
	 	\tilde E_x =\frac{1}{\sigma_0}[j_x-\frac{\omega\tau_e}{B} (B_z j_y-B_y j_z)], \qquad\qquad\\
	 	\label{8jt}	
	 	\tilde	E_y =\frac{1}{\sigma_0}[j_y-\frac{\omega\tau_e}{B} (B_x j_z-B_z j_x)], \qquad\qquad \\
	 	\label{9jt}	
	 	\tilde  E_z =\frac{1}{\sigma_0}[j_z-\frac{\omega\tau_e}{B} (B_y j_x-B_x j_y)]. \qquad\qquad 
	 \end{eqnarray}
	 
	 With account of (\ref{2m}), (\ref{2j}), this solution is written in the vector form as:
	 
	 \begin{eqnarray}
	 	\label{10jt}
	 	{\bf \tilde E }={\bf E'+E_{ex}}+\frac{\lambda_0}{\sigma_0}{\bf\nabla}T= \frac{\bf j}{\sigma_0}-\frac{\omega\tau_e}{B\sigma_0}
	 	{\bf[B\times j]}\\ = \frac{c}{4\pi \sigma_0}{\bf \nabla \times B}-
	 	\frac{c\, \omega\tau_e}{4\pi\sigma_0 B}{\bf B\times[\nabla\times B]} \nonumber.
	 \end{eqnarray}  
	 
	 \begin{eqnarray}
	 	\label{11jt}
	 	{\bf E }=-\frac{1}{c}{\bf [v\times B]}-{\bf E_{ex} }-\frac{\lambda_0}{\sigma_0}{\bf\nabla}T+
	 	\frac{c}{4\pi \sigma_0}{\bf \nabla \times B} \\
	 	\label{12jt}
	 	-\frac{c\, \omega\tau_e }{4\pi\sigma_0 B}{\bf B\times[\nabla\times B]}-\frac{{\bf\nabla} P_e}{e n_e}. \nonumber
	 \end{eqnarray}
	 
	 In presence of {thermal diffusion} and  external electric field $\bf E_{ex} $  we obtain, after applying the curl procedure to the equation \eqref{11jt}, and using \eqref{1m}, the following equation  at $P_e=n_e kT$: 
	 
	 \begin{eqnarray}
	 	\label{4fa}
	 	\frac{\partial{\bf B}}{\partial t}={\bf \nabla}\times 
	 	[{\bf v\times B}] + c{\bf \nabla\times E_{ex}} 
	 	+c\nabla \biggl(\frac{\lambda_0}{\sigma_0}\biggr)\times{\nabla T} \qquad \qquad\\ \nonumber-\frac{ck}{e n_e}[{\bf \nabla}n_e\times {\bf \nabla} T]
	 	+\frac{c^2}{4\pi\sigma_0}{\bf \nabla^2  B}
	 	-\frac{c^2}{4\pi}[{\bf \nabla}(\frac{1}{\sigma_0})\times 
	 	{\bf\nabla\times B}]
	 	\\ \nonumber +\frac{c^2}{4\pi} {\bf\nabla}\times
	 	\big[\frac{\omega\tau_e}{\sigma_0 B}{\bf B}\times[\nabla\times{\bf B}]\big].\nonumber
	 \end{eqnarray} 
\subsection{The Biermann battery}

{The third and fourth terms on the right side of  the Eq.\eqref{4fa}
	may create a seed magnetic field in the 
	unmagnetized media. The third term is related to the 
	{thermal diffusion}, and the fourth one is connected 
	with a current produces by a pressure gradient 
	(baro-diffusion). }The seed magnetic field is created 
when the surfaces of the constant concentration 
$n_e$ don't coincide with surfaces of constant 
temperature, 
$$\nabla n_e \times \nabla T \neq 0.$$
These two terms are representing the Biermann battery \cite{bier1,bier2}.

The Eq.\eqref{4fa} is valid for {varying values} of the parameters $\lambda_0$ and $\sigma_0$.
At constant value $\lambda_0/\sigma_0=3k/2e=const$, in the fully ionized non-degenerate Lorenz  plasma,  the first temperature term is zero. At in this conditions  {thermal diffusion} is not participating  in creation of the seed magnetic field. When this term is variable, 
$$\lambda_0/\sigma_0=f(\rho,T)$$
the seed magnetic field is created by a {thermal diffusion} at non-zero value of
$$\frac{\partial f}{\partial\rho}[\nabla\rho \times \nabla T]$$.  

In neutron star crust the electrons are strongly degenerated, and the conditions of the seed field creation are changed. 

Let us consider  the case of strongly degenerate, non-relativistic electrons.
In this case the coefficients of conductivity and {thermal diffusion}, calculated in \cite{llsp}, have the form

\begin{eqnarray}
	\label{4fb}
	\sigma_0^d=
	\frac{e^2 n_e}{m_e}\tau_d,
	\qquad \lambda_0^d=
	\frac{4\pi^3 e}{(9\pi)^{1/3}}\frac{n_e^{1/3}k^2 T}{h^2} \tau_d.
\end{eqnarray} 
Here
\begin{eqnarray}
	\label{4fc}
	\tau_d=\frac{3}{32\pi^2}\frac{h^3 n_e}{Z^2e^4 m_e 
		n_N\Lambda}, \qquad h\,\,\,\,{\mbox{ is the
			Planck constant}}.
\end{eqnarray} 
The ratio of transport coefficients is written as

\begin{eqnarray}
	\label{4fd}
	\frac{\lambda_0^d}{\sigma_0^d}=\frac{k}{e}\, 
	\frac{4\pi^3}{(9\pi)^{1/3}}\frac{k T m_e}{h^2  n_e^{2/3}}
\end{eqnarray}    
what may give a non-zero value 
$\nabla n_e \times \nabla T \neq 0.$

For strongly degenerate plasma 
the baro-diffusion term in \eqref{4fa} should be taken, using \eqref{2j}, with the equation of state 
in the form (see \cite{llsp})

\begin{eqnarray}
	\label{4fe}
	P_e=2kT \big(\frac{2kT m_e}{h^2}\big)^{3/2} G_{5/2}(x_0), \\ \nonumber n_e=2\big(\frac{2kT m_e}{h^2}\big)^{3/2} G_{3/2}(x_0). 
\end{eqnarray}  
We obtain from 
\cite{llsp, BK2001}

\begin{eqnarray}
	\label{4fe1}
	x_0=\frac{\varepsilon_{fe}}{kT}=
	\frac{(3\pi^2 n_e)^{2/3}h^2}{8\pi^2m_ekT}\gg 1,
	\qquad\qquad\\
	G_{5/2}(x_0)\approx \frac{x_0^{5/2}}{\Gamma(7/2)}
	+\frac{\pi^2}{4}\frac{x_0^{1/2}}{\Gamma(5/2)}, \qquad \quad\\ \nonumber
	G_{3/2}(x_0)\approx \frac{x_0^{3/2}}{\Gamma(5/2)}
	+\frac{\pi^2}{12}\,\frac{x_0^{-1/2}}{\Gamma(3/2)},
	\qquad \quad \nonumber\\
	\label{4fe2}
	P_e=\frac{(3\pi^2)^{2/3}}
	{20\pi^2}\frac{h^2 n_e^{5/3}}{m_e}\big(1+\frac{5\pi^2}{8 x_0^2}\big)=P_{ed}\big(1+\frac{5\pi^2}{8 x_0^2}\big).
\end{eqnarray}  
The creation of the seed magnetic field in the 
strongly degenerate plasma is described by the equation

\begin{eqnarray}\label{4ff}
	\frac{\partial{\bf B}}{\partial t}=c\nabla \biggl(\frac{\lambda_0^d}{\sigma_0^d}\biggr)\times{\nabla T}
	+\frac{c}{e}\, 
	\nabla\times\biggl( \frac{{\bf\nabla}P_e}{n_e}\biggr)\\
\end{eqnarray}
We obtain from \eqref{4fd} and \eqref{4fe2}

\begin{eqnarray}\label{4fg}
	\nabla \biggl(\frac{\lambda_0^d}{\sigma_0^d}\biggr)
	=\frac{k}{e}\, 
	\frac{4\pi^3}{(9\pi)^{1/3}}\frac{k m_e}{h^2 }
	\biggl(\frac{\nabla T}{n_e^{2/3}}- \frac{2}{3}\frac{T\nabla n_e}{n_e^{5/3}}\biggr), \qquad \qquad\\
	\label{g2}
	\nabla\times\biggl(\frac{{\bf\nabla}P_e}{n_e}\biggr)\,=-\frac{\nabla n_e}{n_e^2}\times {\bf\nabla P_e}= -2\frac{P_{ed}}{n_e^2}\frac{5\pi^2}{8 x_0^2 T}
	\nabla n_e\times{\nabla T}.\quad
\end{eqnarray}  

With account of \eqref{4fg},\eqref{g2} we have from \eqref{4ff}

\begin{eqnarray}\label{4fh}
	\frac{\partial{\bf B}}{\partial t}=-\biggl[   
	\frac{4\pi^{8/3}}{3^{2/3}} \frac{c}{e}\frac{k^2Tm_e }{h^2 \,n_e^{5/3}} +
	\frac{8\,\pi^{7/3}}{3^{5/3}}\,
	\frac{c}{e}\,\frac{k^2Tm_e}{h^2n_e^{5/3}} \biggr]
	(\nabla n_e\times{\nabla T})\quad
\end{eqnarray}

Using \eqref{4fe2}, this equation is rewritten as
(see also \cite{Blandford-Applegate-Hernquist})

\begin{eqnarray}
	\label{4fh1}
	\frac{\partial{\bf B}}{\partial t}=
	-\biggl[40.7109+
	18.5312\biggr]
	\frac{c}{e}
	\frac{k^2 T m_e}{h^2\,n_e^{5/3}}
	(\nabla n_e\times{\nabla T})
\end{eqnarray}

In \eqref{4fh1} the left term in the brackets is related to baro-diffusion, and the right one is for thermal diffusion. The input of both terms here is comparable, with a constant ratio.

\subsection{The system of MHD equations related to magnetic field behavior}

   In the system with temperature gradients, 
   external electric field ${\bf E_{ex}}$ and external magnetic field ${\bf B_{ex}}$, connected with external electric current ${\bf j_{ex}}$, there are {self-fields} ${\bf E_{in}}$, ${\bf B_{in}}$ and {self} electric current ${\bf j_{in}}$. The {self-fields} are connected by the first Maxwell equation (\ref{1m}).  
   Finally, we obtain the following system of equation for determination of these values:
   
   \begin{eqnarray}
   	\label{1ex}
   	\frac{1}{c}\frac{{\partial\bf B}_{in}}{\partial t}=-{\bf \nabla\,
   		\times\,E}, \quad {\bf\nabla \cdot B}_{in} =0,\quad {\bf j_{in}}=\frac{c}{4\pi} {\bf\nabla\,\times\,B}_{in},	\qquad	
   	\\
   	\label{2ex}	
   	{\bf E}=-\frac{1}{c}{\bf [v\times B]}-	
   	{\bf E_{ex} }-\frac{\lambda_0}{\sigma_0}{\bf\nabla}T+
   	\frac{c}{4\pi \sigma_0}{\bf \nabla \times B} \qquad \nonumber\\
   	-\frac{c\,\omega\tau_e}{4\pi\sigma_0 B}
   	{\bf B\times[\nabla \times B]}-\frac{{\bf\nabla} P_e}{e n_e},\quad \qquad\\ 	
   	\label{3ex}
   	{\bf j}_{in}=\frac{\sigma_0}{1+\omega^2 {\tau_e}^2}
   	\biggl[{\bf E'}+\frac{\omega^2 {\tau_e}^2}{B^2}{\bf (B\cdot E')B} -\frac{\omega\tau_e}{B} \bf{[B\times E']}\biggr],\quad\\
   	{\bf E'}={\bf E}+\frac{1}{c}{{\bf [v\times B]}}+\frac{{\nabla} P_e}{e n_e},\,\,\quad\qquad 	\\		
   	\label{4ex}
   	\frac{c}{4\pi} {\bf\nabla\,\times\,B}_{ex}^v={\bf j_{ex}},\quad 
   	{\bf\nabla \cdot B}_{ex}=0,\quad
   	{\bf B}_{ex}={\bf B}_{ex}^v+{\bf B}_{ex}^{vf},\qquad \\
   	\label{5ex}		
   	{\bf j}_{ex}=\frac{\lambda_0}{1+\omega^2 {\tau_e}^2}
   	\biggl[{\bf\nabla} T+\frac{\omega^2 {\tau_e}^2}{B^2}{\bf (B\cdot  \nabla}T){\bf B} -\frac{\omega\tau_e}{B} {\bf[B\times \nabla}T]\biggr]\quad\\ \nonumber
   	+\frac{\sigma_0}{1+\omega^2 {\tau_e}^2}
   	\biggl[{\bf E_{ex}}+\frac{\omega^2 {\tau_e}^2}{B^2}{\bf (B\cdot E_{ex})B} -\frac{\omega\tau_e}{B} \bf{[B\times E_{ex}]}\biggr],\qquad\\		
   	\label{6ex} 		
   	{\bf j}= {\bf j_{in}}+{\bf j_{ex}},  \qquad 
   	{\bf B}={\bf B}_{in}+{\bf B}_{ex}. \qquad  \qquad\qquad\qquad  
   \end{eqnarray}  
   
   Here ${\bf E}$, ${\bf E'}$ are inner electric fields measured in the laboratory, and in the co-moving coordinate system,  respectively. In addition there are given external
   {electric} field ${\bf E_{ex}}$, external current ${\bf
   	j_{ex}}$, creating a vortex external magnetic field $
   {\bf {B_{ex}^v}}$; and independent vortex-free magnetic  field ${\bf {B_{ex}^{vf}}}$.

   There is a complicate situation due to presence of a external fields, currents and  thermal diffusion.
   All currents are producing  magnetic fields, which,  together with the external field ${\bf {B_{ex}^{vf}}}$, are taking part in the formation of  Hall currents.	
   The currents are produced in different processes, but the magnetic field which they have created, is acting together, as one vector ${\bf B}$. These properties are represented in the system of equations (\ref{1ex})-(\ref{6ex}).

\section{    Several applications }

    \subsection{Thrusters}
    
Thrusters are reactive engines at low power, and very high efficiency, based on the electromagnetic acceleration of plasma. A conception of this engine was suggested in the year 1957 by A.I. Morozov \cite{moroz1957}. A simplified version of their action is based on the consideration of the ideal plasma behavior inside the mutually perpendicular external electric ${\bf E_{ex}}$ and magnetic ${\bf B_{ex}}$ fields, where the {self electric} field  in the laboratory system ${\bf E}$ is zero. For the ideal plasma with infinite conductivity $\sigma_0$ we have from (\ref{2ex}) the relation, determining the equilibrium plasma velocity {\bf v} in these conditions 

\begin{eqnarray}
	\label{tr1}
	{\bf E} = -\frac{1}{c}{\bf [v\times B]} - {\bf E_{ex} }=0.
\end{eqnarray}

This equation has an exact solution 

\begin{eqnarray}
	\label{tr2}
	{\bf v} = c\frac{\bf E_{ex}\times B} {B^2}.  
\end{eqnarray}
As followed from (\ref{1ex}), at zero  ${\bf E}$ the inner magnetic field  ${\bf B_{in}}$=0
therefore, we take here ${\bf B}={\bf B_{ex}}$.

The review's devoted to the 30th anniversary of small plasma thrusters (SPT) operation in space
are given in Refs. \cite{moroz2003} (about the conceptual development of stationary plasma thrusters), and \cite{koz2003} (about stationary plasma thrusters operate in space).

Usually we have situations when $E\ll B$. so that the speed of the {outflowing} plasma $v \ll c$. If, by some reason, the value of ${\bf E}$ is approaching that of  ${\bf B}$, the speed of the {outflowing} plasma is approaching {speed of light} $c$. In this situation the approximate Lorentz transformation formulae should be substituted by its exact version \cite{llft}, and instead of \eqref{tr1}, \eqref{tr2} we  obtain following relations for the geometry where, in Cartesian coordinates $(x,y,z)$

\begin{eqnarray}
	\label{tr3}
	{\bf v} = (V,0,0), \quad {\bf E} = (0,E,0), \quad {\bf B} = (0,0,B),
\end{eqnarray}
the following result

\begin{eqnarray}
	\label{tr4}
	E = -\frac{V}{c}\frac{B}{\sqrt{1-V^2/c^2}} - E_{ex} =0.
\end{eqnarray}

This equation has an exact solution 

\begin{eqnarray}
	\label{tr5}
	V = {c}\frac{E_{ex}}{\sqrt{B_{ex}^2+ {E_{ex}}^2}}. 
\end{eqnarray}

\subsection{Neutron star crust}	

    Biermann battery effects connected with a magnetic field behavior, and its possible role in the seed magnetic field creation in the neutron star (NS) crust, have been  considered in  \cite{Blandford-Applegate-Hernquist}. 
   The authors used transport coefficients from the works 
   \cite{fi76, fi79, fi81} where numerical estimations of some integral have been made. 
   
   {Here we obtain a fully analytic solution of this problem, with approximate treating of relativistic effects.}

   In the middle aged NS crust a temperature is moderate, and a density is very high, so the electrons are strongly degenerate and ultrarelativistic. {We use Eq. \eqref{4ff} with relativistic presentation for $P_e$ and effective electron mass $m_*$, instead of 
   	$m_e$: }
   
   \begin{eqnarray}
   	\label{nc}
   	m_* \approx {m_e}{\sqrt{1+\frac{p_{fe}^2}{(m_e c)^2}}},
   	\qquad p_{fe}=\bigg(\frac{3n_e}{8\pi}\biggr)^{1/3}h. 
   \end{eqnarray}
   
   In this case the coefficients of conductivity and thermal diffusion, with account of \eqref{nc}, are taken in the form
   
   \begin{eqnarray}
   	\label{nc1}
   	\sigma_0^{du}=
   	\frac{e^2 n_e}{m_*}\tau_{du},
   	\qquad \lambda_0^{du}=
   	\frac{4\pi^3 e}{(9\pi)^{1/3}}\frac{n_e^{1/3}k^2 T}{h^2} \tau_{du},
   \end{eqnarray} 
   with
   \begin{eqnarray}
   	\label{nc2}
   	\tau_{du}=\frac{3}{32\pi^2}\frac{h^3 n_e}{Z^2e^4 m_* 
   		n_N\Lambda}, \qquad h\,\,\,\,{\mbox{ is the
   			Planck constant}}.
   \end{eqnarray} 
   The ratio of transport coefficients is written as
   
   \begin{eqnarray}
   	\label{nc3}
   	\frac{\lambda_0^{du}}{\sigma_0^{du}}=\frac{k}{e}\, 
   	\frac{4\pi^{8/3}}{3^{2/3}}\frac{k T m_*}{h^2  n_e^{2/3}},
   \end{eqnarray}    
   with possible non-zero value of
   $\nabla n_e \times \nabla T \neq 0.$
   
   For
   the baro-diffusion term in \eqref{2j},  \eqref{12jt} of degenerate  electrons the equation of state in the relativistic limit is written 
   in the form (see \cite{BK2001})

   \begin{eqnarray}
   	\label{nc6}
   	P_{d}=\frac{1}{8}\biggl(\frac{3}{\pi}\biggr)^{1/3} 
   	c\, h\, n_e^{4/3}\biggl[1+\frac{2\pi^{2/3}}{3^{5/3}}
   	\biggl(\frac{2\pi kT}{h\,c}\biggr)^2 n_e^{-2/3} \biggr]
   \end{eqnarray}  
   The creation of the seed magnetic field in the 
   strongly degenerate plasma is described by the equation \eqref{4ff}
   We obtain from \eqref{nc3} and \eqref{nc6}
   
   \begin{eqnarray}
   	\label{nc8}
   	\nabla \biggl(\frac{\lambda_0^d}{\sigma_0^d}\biggr)
   	=\frac{k}{e}\, 
   	\frac{4\pi^{8/3}}{3^{2/3}}\frac{k m_*}{h^2 }
   	\biggl(\frac{\nabla T}{n_e^{2/3}}- \frac{2}{3}\frac{T\nabla n_e}{n_e^{5/3}}\biggr),
   \end{eqnarray}
   
   \begin{eqnarray}
   	\label{nc9}
   	\nabla\times\biggl(\frac{\nabla P_{d}}{n_e}\biggr)=-\frac{\nabla n_e}{n_e^2}\times 
   	{\nabla P_{d}} \qquad \qquad \qquad \\ \nonumber
   	=-\frac{\pi^{1/3} ch}{2 \cdot 3^{4/3}n_e^{4/3}T} \biggl(\frac{2\pi kT}{h\,c}\biggr)^2 (\nabla n_e\times{\nabla T}).
   \end{eqnarray}

   With account of \eqref{nc8},\eqref{nc9} we have from \eqref{4ff}
   
   \begin{eqnarray}
   	\label{nc10}
   	\frac{\partial{\bf B}}{\partial t}=-\biggl[
   	\frac{ck}{e}\, 
   	\frac{    8\pi^{8/3}}{3^{5/3}}\frac{kT m_*}{h^2 n_e^{5/3}}\qquad \qquad \qquad\\ \nonumber
   	+\frac{c}{e}\,
   	\frac{\pi^{1/3}ch}{2 \cdot3^{4/3}n_e^{4/3}T} \biggl(\frac{2\pi kT}{h\,c}\biggr)^2\biggr] 
   	(\nabla n_e\times{\nabla T})
   \end{eqnarray}

   This equation is rewritten as
   (see also \cite{Blandford-Applegate-Hernquist})

   \begin{eqnarray}\label{nc11}
   	\frac{\partial{\bf B}}{\partial t}=-\frac{ck}{e}\,
   	\frac{2\pi^{7/3}}{3^{4/3}n_e^{4/3}} \biggl(\frac{kT}{h\,c}\biggr) 
   	\biggl(1+\frac{4\pi^{1/3}}{ 3^{1/3}}\frac{m_*c}{h n_e^{1/3}}  \biggr)
   	(\nabla n_e\times{\nabla T}). \quad
   \end{eqnarray}
   
   In this consideration the baro-diffusion term is based on exact equation of state for relativistic electrons, while in the thermal diffusion term we used corrected non-relativistic kinetic coefficients, similar to \cite{fi76}. In this situation the physical interpretation of terms in the brackets of \eqref{nc11} is not so clear.

	\subsection{Magnetized plasma cylinder}

    The model of the magnetized {plasma cylinder} with a radial temperature gradient was investigated in \cite{bkg2}. The temperature gradient created the  {thermal diffusion electric} current, in the external uniform magnetic field along the cylinder axis (see Fig.1). 
    
    It was obtained in \cite{bkg2}, that thermal diffusional Hall currents in this situation  create magnetic field of the sign, opposite the the external one, and can substantially decrease the resulting magnetic field. This behavior is in accordance with the Lenz's law \cite{tamm} in electrodynamic.

    \begin{figure}
    	\begin{center}
    	\includegraphics[width=0.42\textwidth]{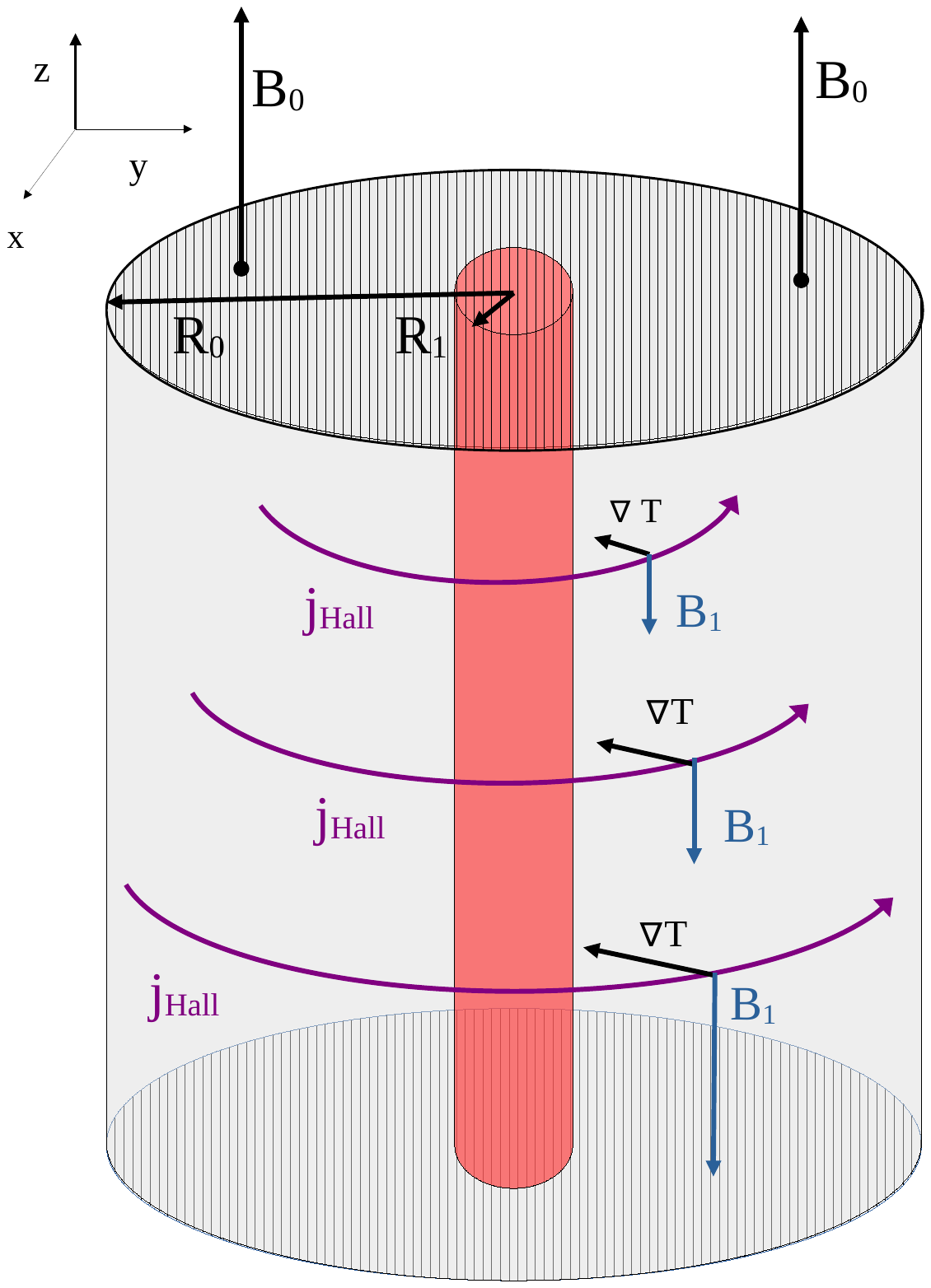}
    	\end{center}	
    	\caption{Conducting cylinder with Hall current $j_{Hall}$, depending on the magnitude of the radial temperature gradient, and external constant magnetic field $B_0$ along its axis. The induced magnetic field $B_1$ is determined by the Hall current. $R_1$ is the radius of the  central heated region with constant temperature $T_0$. {Torus} region, colored in gray, contains Hall current and associated magnetic field, which has an opposite direction to the external field $B_0$, decreasing the resulting field along the cylinder. }\label{cylinder1}
    \end{figure}

\subsection{Magnetized plasma torus}
	
The plasma torus configurations are used in construction of devices for thermonuclear energy production (Tokamak, Stellarator) 
\cite{sol1972, ksh}.
Building of ITER (International Thermonuclear Experimental Reactor) is based on the Tokamak installation \cite{wikit}.

In astrophysics the plasma torus may be formed around compact objects (white dwarfs, neutron stars, black holes) during accretion in the binary system.
The giant magnetized plasma torus configuration, formed around the central nuclei in the active galactic nuclei and quasars, are considered now as a most plausible model, explaining many different observational features \cite{lr1998, aar2015}.

\begin{figure}
			\begin{center} 
	 \includegraphics[width=0.60\textwidth]{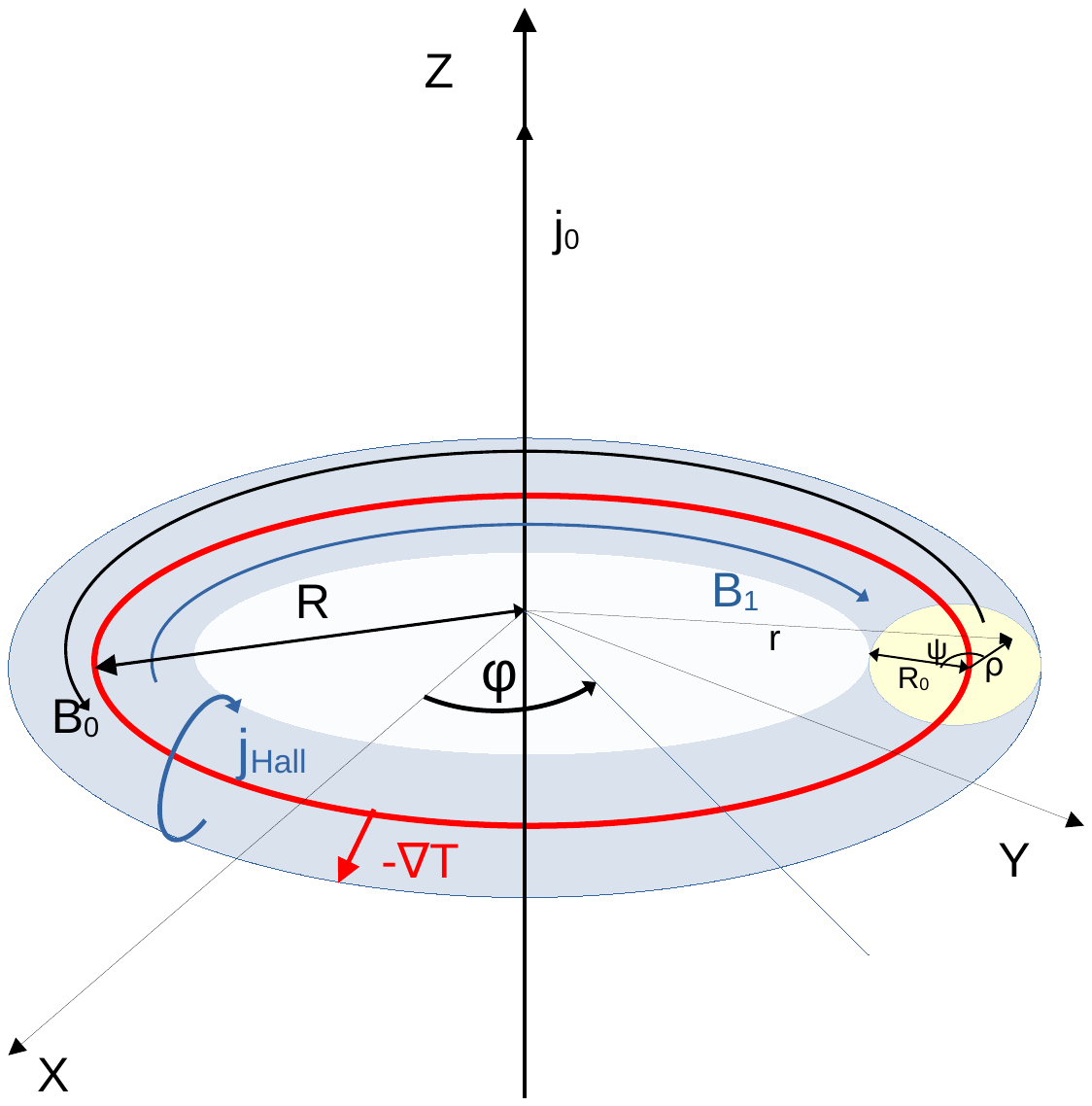}
			\end{center}	
			\caption{Torus  with a initial electric current $j_0$, that produce circular magnetic field $B_0$. $B_0$  and temperature gradient $\nabla T$ create Hall electric current $j_{Hall}$.  The induced magnetic field $B_1$ is determined by the Hall current $j_{Hall}$ and is oriented in the opposite direction to the $B_{0}$.}\label{torus}
		\end{figure}

Here we  consider the structure of a circular plasma torus, using Eqs. (\ref{1ex}) -(\ref{6ex}). 
Through the center of the torus there is a linear {electric} current, situated perpendicular to the symmetry plain of the torus $(x,y)$, with a linear current  ${J_z^{ex}}$,  see Fig.\ref{torus}. The cross section of the torus, perpendicular to  $z$-axis, is supposed to be circular. In the symmetry plane $z=0$ the torus occupies the space between the radius $R_{out}$ and $R_{in}$. 
The cross section of the torus along $z$-axis is also taken as a circular, with the radius $R_t$, so that
$$R_{out}=R_{in}+2R_t.$$

\noindent Using Biot–Savart law, the magnetic field of the infinite linear current is represented by concentric circles around the current in the plane perpendicular to it. This external  field has only $\phi$ component and is written as \cite{tamm}:

\begin{equation}\label{BJ}
	B_\phi^{ex}=	\frac{2J_z^{ex}}{cr}.	
\end{equation}

The torus surface is defined by the 4-th order equation. It is formed by a rotation of a circle with the radius $R_t$,
around a $z$-axis, when the center of the circle has the radius $R=R_{in}+R_t$. The equation of this surface is written as follows

\begin{eqnarray}
	(\sqrt{x^2+y^2}-R)^2+z^2=R_t^2, \quad  {\rm which\,\,is \,\, reduced\,\,to}\quad \nonumber\\(x^2+y^2+z^2+R^2-R_t^2)^2=4R^2(x^2+y^2).\,\,\,\quad \label{tor}
\end{eqnarray}	
By heating the central ring of the torus and formation of the temperature gradient ${ \nabla T}$,  a heat flux is created inside the torus along its smaller radius $\rho$, directed to the surface of the torus.
In the {cylindrical coordinate system} $r,\phi,z$ the  the coordinate  $\rho$ is defined as: 

\begin{equation}\label{BJ1}
	\rho^2=(\sqrt{x^2+y^2}-R)^2+z^2=(r-R)^2+z^2,\quad  
	0\leq\rho\leq R_t,
\end{equation}
so that in the plane $z=0$, \quad $\rho d\rho=(r-R)dr$, for $R_{in}<r<R$.
For a constant heat flux $Q$ through the unit length of the large circle of the torus, the temperature gradient along the small torus radius $\rho$ is determined as

\begin{equation}\label{BJ3}
	q=\frac{Q}{2\pi \rho}=-\frac{\kappa_0}{1+ (\omega \tau_e)^2}
	\frac{dT}{d\rho}, \quad \kappa_0=\frac{320}{3\pi}
	\frac{k^2 T n_e \tau_e}{m_e}.
\end{equation}
Here the scalar heat conductivity coefficient $\kappa_0$
is determined in \cite{BK2001}.  The {thermal diffusion} is forming the {electric} current along the temperature gradient along the $\rho$ coordinate $j_{in,\rho}$, and a hall current $j_{in,\psi}$ along the small circle of the torus:

\begin{equation}
	\label{BJ4}
	j_{in,\rho} =  \frac{\lambda_0}{1+ (\omega \tau_e)^2} \frac{d T}{d \rho},
	\qquad
	j_{in,\psi} = \lambda_0\frac{ \omega \tau_e}{1+ (\omega \tau_e)^2} \frac{d T}{d\rho}.
\end{equation}
Due to the axial symmetry, the magnetic field inside the torus is produced only by the Hall component $j_{in,\psi}$,corresponding to a circular current along the small circle of the torus, see Fig.\ref{torus}. 

{In astrophysical conditions, the current along the $\rho$ coordinate is  crossing freely the torus boundary. The surrounding background plasma is supplying the electron flux inside the torus, so that its total electrical charge is remaining zero.
	The torus is used in the construction of the devices for obtaining controlled thermonuclear reaction \cite{ksh}, where these currents are concentrated in the wire wound up in the torus around its small radius. Such current is creating the toroidal field along the big circle the torus. In our model the external toroidal magnetic field is created by the central linear current.	}

To calculate the magnetic field created by the {thermal diffusion} $\psi$ - Hall current component, we use the way similar to the one, which used for  calculation of {torus} field created by the current through  the wound up wire. If $I$ is the current along the wire, and $n$ is the number of wire small circles on the unit length of the torus big circle, than the external circular magnetic field $B_\varphi(r)$ inside  the torus, at $R_t\ll R$, is approximately written as  \cite{torB} 

\begin{equation}
	\label{BJ5}
	B_\varphi^{in}(r)=\frac{4 \pi}{c} n I \frac{r}{R}, \quad r\approx R.
\end{equation}

In the plasma torus there is a circular Hall current
density $j_{in,\psi}$, defined in \eqref{BJ4}.
The magnetic field formed by the Hall current could be considered as sum of thin circular currents, similar to the wire current in the laboratory torus, considered above. 
The $\psi$ - current in the torus on the radial distance $r$ from the center is equal to 
($j_{in,\psi} 2\pi r d\rho$).  The magnetic field circulation produced by  this current along the large circle of the torus,
is approximately equal to $-2\pi R \,dB_\varphi^{in}(r)$. So the magnetic field $dB_\phi^{in}$ produced by this sheet of current at the radius $r$ is written as:

\begin{equation}
	\label{BJ6}
	dB_\varphi(r)= -\frac{4 \pi}{c} j_{in,\psi} \frac{r}{R}d\rho, 
	\quad R_{in}<r<R.
\end{equation}

It follows from \eqref{BJ5}, after substituting ($nI$) by  ($j_{ex,\psi}(\rho)d\rho$). The Hall  magnetic  field from the {total} current in the plasma torus is obtained by integration of \eqref{BJ6}  over the torus thickness, as
\begin{eqnarray}
	\label{BJ7}
	B_\varphi(r)=-\frac{4 \pi}{c}\int_{R-r}^{R-R_{in}} j_{in,\psi} \frac{r}{R}d\rho, =-\frac{4 \pi}{Rc}\int_{R-r}^{R-R_{in}} j_{in,\psi} r d\rho, \quad\\ \nonumber R_{in}<r<R.
\end{eqnarray}
On the symmetry plane $z=0$ there is a connection 
$\rho=R-r$, $d\rho=-dr$. For a constant current density 
we have after integration:

\begin{eqnarray}
	\label{BJ8}
	B_\varphi^{in}(r)=-\frac{4 \pi}{Rc}\int_{R-r}^{R-R_{in}} j_{in,\psi} 
	(R-\rho) d\rho=\\ \nonumber-\frac{2 \pi}{Rc}j_{in,\psi}
	(r^2-R_{in}^2), \quad
	R_{in}<r<R.
\end{eqnarray}

The magnetic field produced by the Hall current has a sign opposite to the field from the central current (Fig.\ref{torus}).
The induced magnetic field is zero at the surface of the torus, and has a maximum (negative) value 

\begin{equation}
	\label{BJ9}
	B_\varphi(R) = -\frac{2 \pi}{c}\frac{j_{in,\psi}}{rc}(R^2-R_{in}^2), \qquad\qquad\qquad
\end{equation}

on the cental circle of the torus.

In numerical  simulation, at {varying coefficients},  we'll use definitions, for the external field of positive sign \eqref{BJ} as $B_\varphi(r) = B_\varphi^{ex}(r)\approx B_0$, and for the {self-field}  \eqref{BJ8} as  $B_\varphi^{in}(r)=B_1(r)$.
We consider a case with a constant heat flux $Q$, through the torus, along its smaller radius $\rho$, and the heat flux through the unit length of the torus $Q_\rho$, which is defined as

\begin{equation}
	\label{BJ10}
	Q_\rho = -2\pi\rho \frac{\kappa_0 }{1+\omega^2\tau_e^2} \frac{dT}{d\rho}. \qquad\qquad
\end{equation}

Here the cyclotron frequency $\omega=\frac{eB}
{m_e c}$ contains $B=B_0+B_1$. 
Equations  (\ref{BJ3}), (\ref{BJ10})  cannot be extended until the axis with $\rho = 0$, because of singularities at zero radius. We presume that the source of a heat is situated along a thin circle along the axis of the torus, 
and is represented by a uniformly heated wire with a radius $\rho_1$, that is much smaller that the small circle radius of torus $\rho_1 << R_t$.
The magnetic field induced by the Hall current is similar to solenoid, having a zero magnetic field at the boundary \cite{tamm}. The equations governing the distribution of the induced magnetic field $B_1$ and temperature $T$ along the $\rho$ coordinate of the torus, are written, in {dimensionless} variables, for the case $R_t\ll R$, similar to the cylinder in \cite{bkg2}, as 

\begin{eqnarray}
	\label{syseqforqmodel}
	\frac{db_1}{dx} = \frac{3C_1 T^{1/2} e Q_\rho\omega_{B0}(1+b_1)}{10 k c B_0 x}, \nonumber\\
	\frac{dT}{dx}=-\frac{1+ C_1^{2} T^{3} \omega_{B0}^2(1+b_1)^2}{2 \pi x C_2 T^{5/2}} Q_\rho.
\end{eqnarray}
This system is solved  at given parameter $Q_\rho$. 
In {dimensionless} parameters with a radius $x=\rho/R_t$, the Hall component $b_1=B_1/ B_0$, and temperature $\tilde T=T/T_0$, the cyclotron frequency is written here as  $\omega_{B} = {eB_{\phi}}/{m_e c} = {e(B_0 + B_1)}/{m_e c} = \omega_{B0} (1+ b_1)$. We have  used also, that ${\lambda_0}/{\kappa_0} = {3 e }/{20 k T}$, 

The constants $C_1$, $C_2$ are determined from relations:
\begin{equation}
	\tau _e = 	\frac{3(kT)^{3/2}}{4Z^2 e^4 n_N \Lambda}\sqrt{\frac{m_e}{2 \pi}} = C_1T^{3/2},
\end{equation}	
\begin{equation}
	\kappa_0 = \frac{40 \sqrt{2} k n_e}{\pi^{3/2}\Lambda n_N}\left( \frac{kT}{e^2 Z}\right)^{2} \left( \frac{kT}{m_e} \right)^{1/2} = C_2 T^{5/2},
\end{equation}
\begin{equation}
	\omega_{B}\tau_e=C_1 T^{3/2}\omega_{B0}(1+b_1).
	\label{eq13a}
\end{equation}
The dimensionless parameters are introduced as

\begin{equation}
	N = \frac{  3 e Q_\rho\omega_{B0} C_1 }
	{10 k c B_0}T_0^{1/2}, \,\,  G = \ C_1^{2} T_0^3 \omega_{B0}^{2},
	\,\, E=\frac{2\pi C_2 T_0^{7/2}}{Q_\rho}.
\end{equation}

\begin{figure}
	\begin{center}
		\includegraphics[width=0.42\textwidth, angle=270]{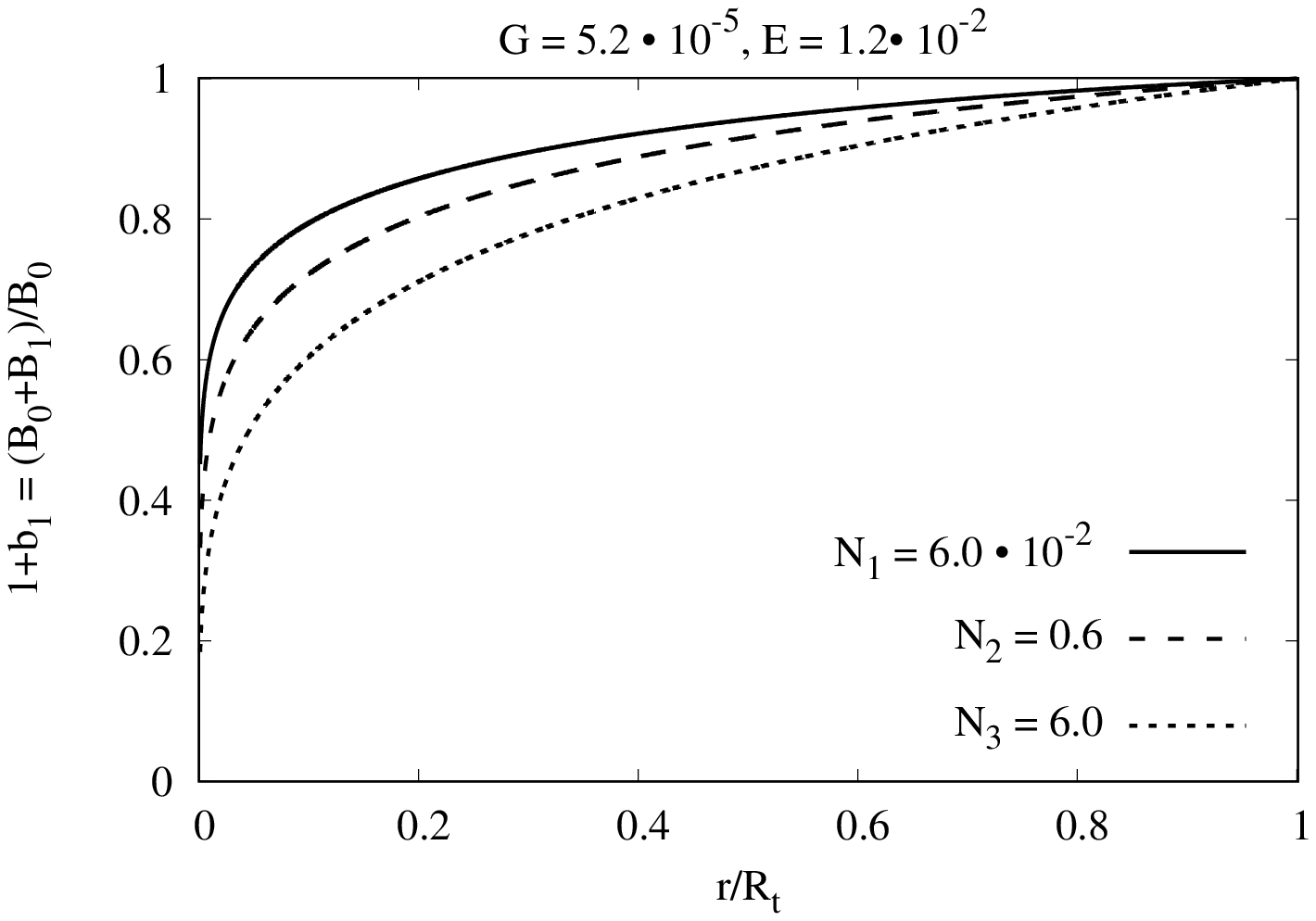}
	\end{center}	
	\caption{
		Magnetic field in the cylinder, induced by the Hall current, for  $G = 5.2\cdot 10^{-5}$, $E = 0.012$, and three {versions} of $N$: $N_1 = 6.0\cdot 10^{-2}$, $N_2 = 0.6$, $N_3 = 6.0$ . These values are related to $Z = 26$, and include combinations:
		\\ $B_{0} = 10^{14}\ G , \quad T_0 = 10^{9} \ K,\quad \rho_0 = 10^{9}$ g/cm$^{3}$ for $N_1$
		\\ $B_{0} = 10^{13}\ G , \quad T_0 = 10^{9} \ K,\quad \rho_0 = 10^{8}$ g/cm$^{3}$ for $N_2$;
		\\ $B_{0} = 10^{12}\ G , \quad T_0 = 10^{9} \ K,\quad \rho_0 = 10^{7}$ g/cm$^{3}$ for $N_3$.}\label{fig3}
\end{figure}

\begin{figure}
	\begin{center}
		\includegraphics[width=0.42\textwidth, angle=270]{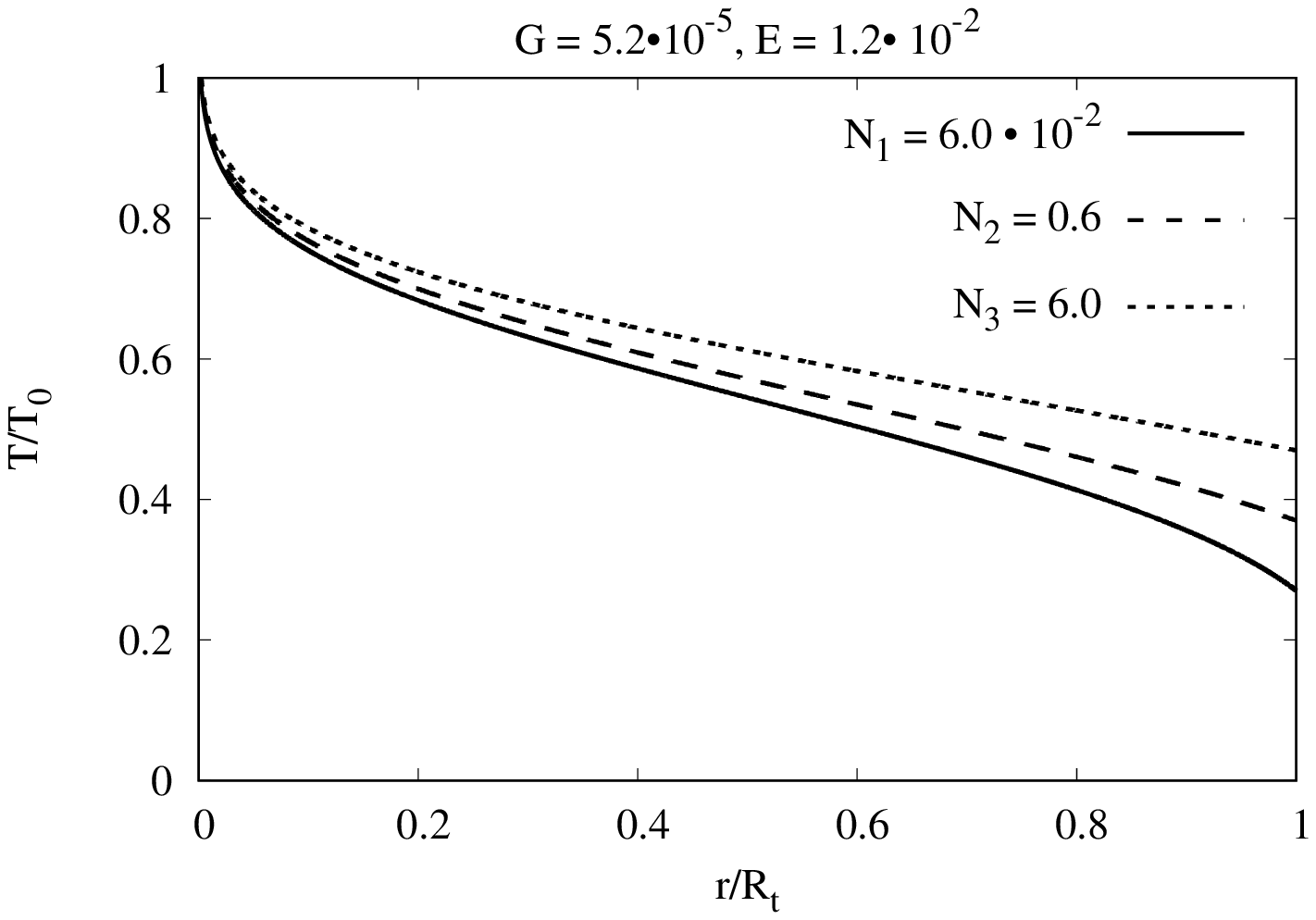}
	\end{center}	
	\caption{Temperature distribution in the torus small circle for the same parameters as in Fig. 3}\label{fig3a}
\end{figure}

\begin{figure}
	\begin{center}
		\includegraphics[width=0.42\textwidth, angle=270]{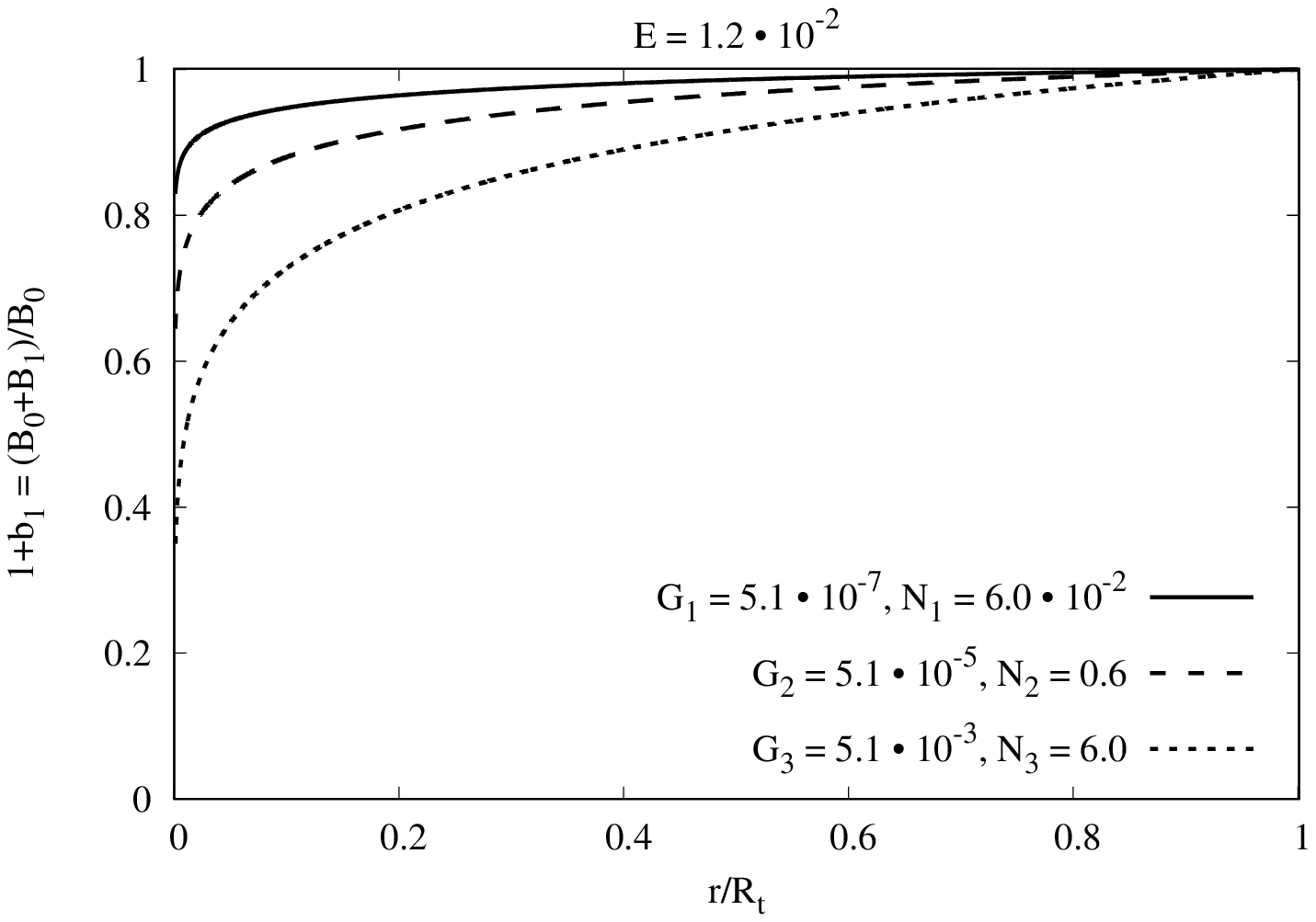}
	\end{center}	
	\caption{
		Magnetic field in the cylinder, induced by the Hall current, for $E=0.012$, and  three {versions}:
		$G_1=5.1\cdot 10^{-7}, N_1 =0.06$.
		$G_2=5.1\cdot 10^{-5},N_2=0.6$;\,\,\,
		$G_3=5.1\cdot 10^{-3},N_3=6.0$;\,\,\,
		These values are related to $Z = 26$, and include combinations
		\\$B_{0} = 10^{13}\, G , \quad T_0 =10^{9}\ K, \quad \rho_0 = 10^{9}$\, g/cm$^{3}$ \quad for $G_1$,$N_1$;
		\\$B_{0} =10^{13}\, G, \quad T_0 =10^{9}\, K, \quad \rho_0 = 10^{8}$\, g/cm$^{3}$ \quad for $G_2$,$N_2$;
		\\$B_{0} = 10^{13}\, G , \quad T_0 = 10^{9}\, K, \quad \rho_0 = 10^{7}$\, g/cm$^{3}$ \quad  for $G_3$, $N_3$.}\label{fig4}
\end{figure}

\begin{figure}
	\begin{center}
		\includegraphics[width=0.42\textwidth, angle=270]{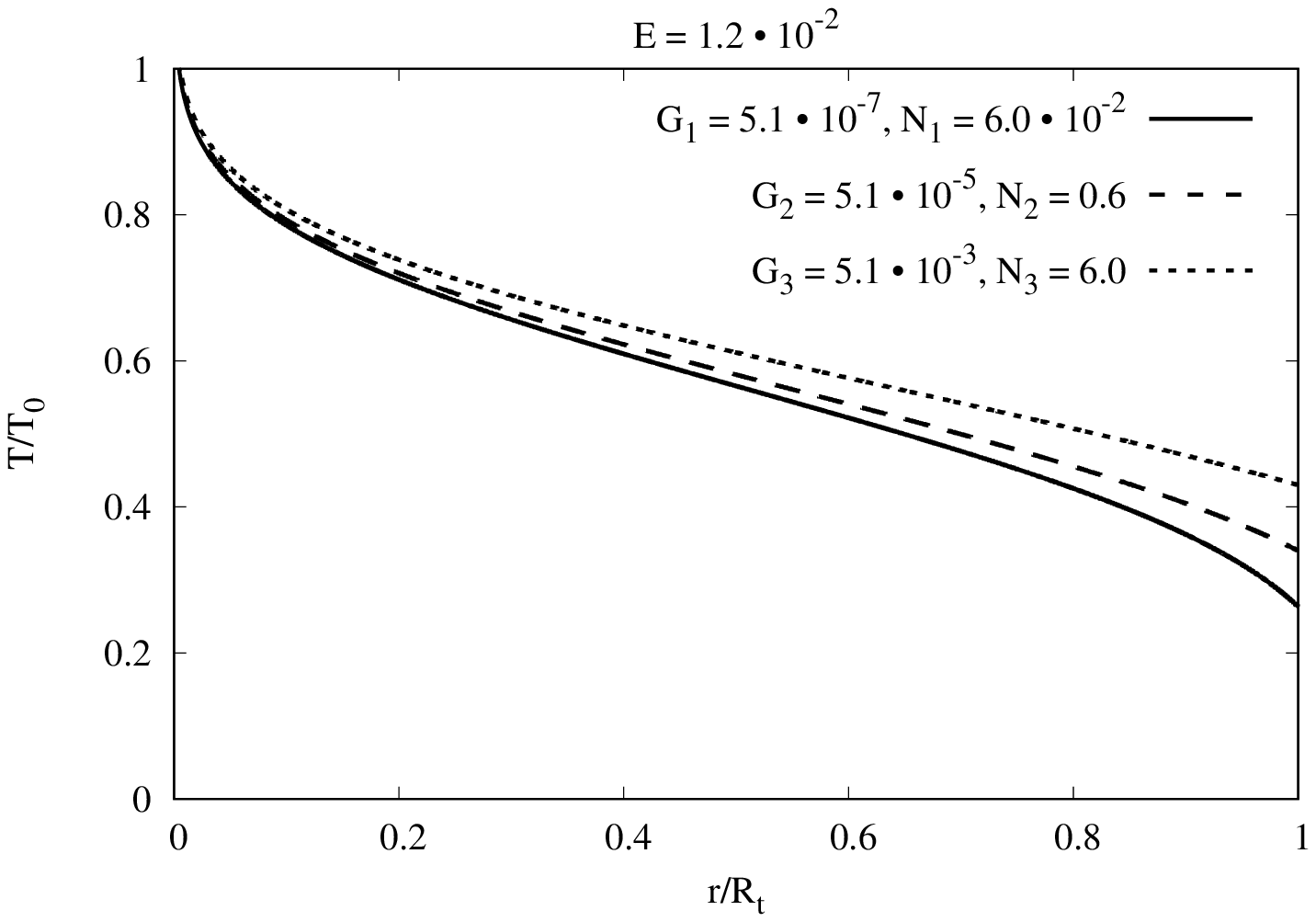}
	\end{center}	
	\caption{Temperature distribution in the torus small circle for the same parameters as in Fig. 5}\label{fig4a}
\end{figure}

\begin{figure}
	\begin{center}
		\includegraphics[width=0.42\textwidth, angle=270]{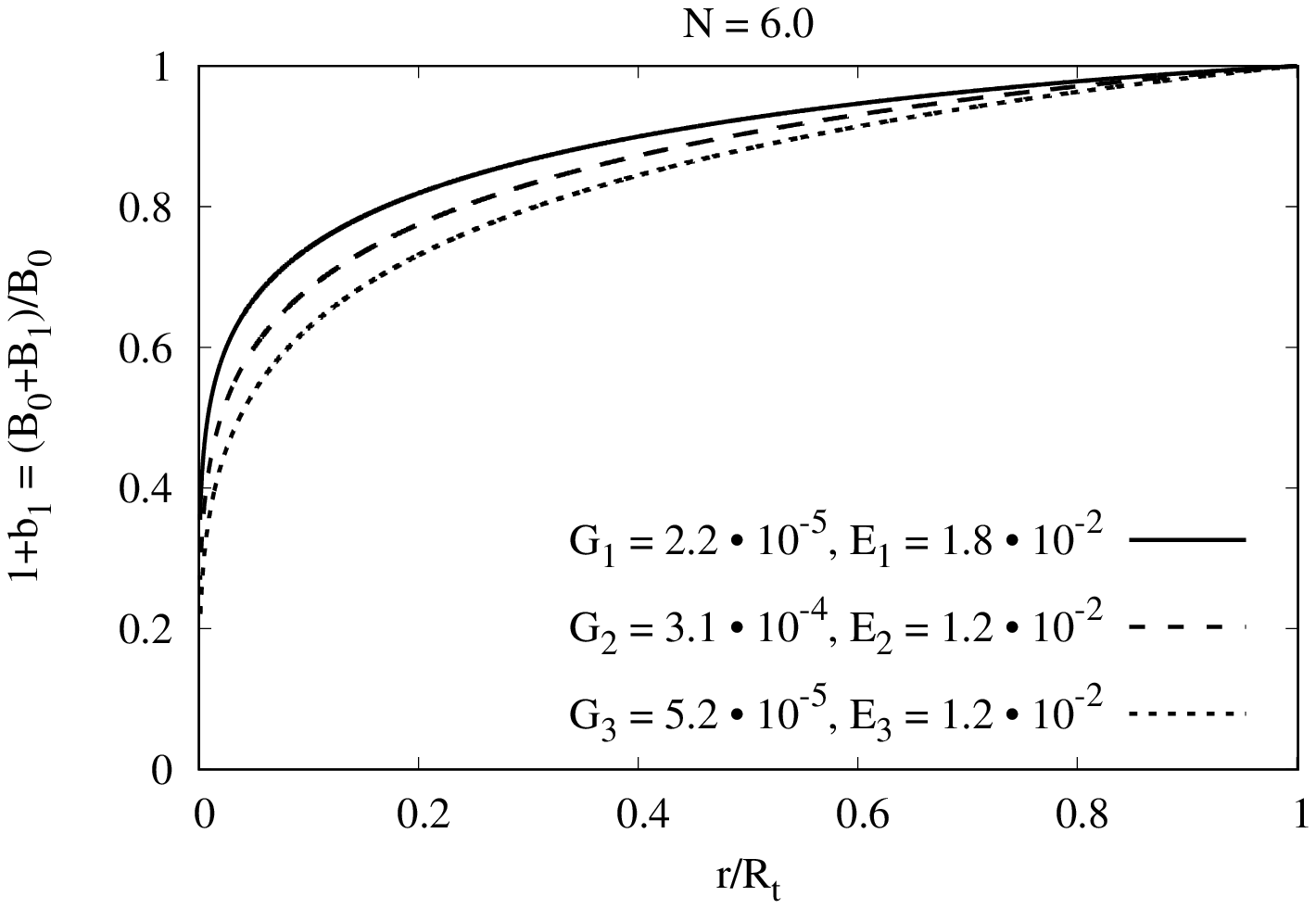}
	\end{center}	
	\caption{Magnetic field in the cylinder, induced by the Hall current, for $N = 6.0$  and three {versions}:
		$G_1 = 2.2\cdot 10^{-5}, E_1 = 0.018$;\,\,\,
		$G_2 = 3.1\cdot 10^{-4}, E_2 = 0.012$;\,\,\,
		$G_3 = 5.2\cdot 10^{-5}, E_3 = 0.012$.
		These values are related to $Z = 26$, and include combinations:
		\\$B_{0} = 10^{13}\ G , \quad T_0 = 3.5\cdot 10^{9} \ K, \quad \rho_0 = 10^{9}$  g/cm$^{3}$\quad for $G_1$,$E_1$;
		\\$B_{0} =10^{13} \ G , \quad T_0 = 1.8\cdot 10^{9} \ K, \quad \rho_0 = 10^{8}$ g/cm$^{3}$\quad for $G_2$,$E_2$;
		\\$B_{0} = 10^{12} \ G ,\quad  T_0 = 10^{9} \ K,\quad \rho_0 = 10^{7}$ g/cm$^{3}$\quad  for $G_3$,$E_3$.}\label{fig5}
\end{figure}

\begin{figure}
	\begin{center}
		\includegraphics[width=0.42\textwidth, angle=270]{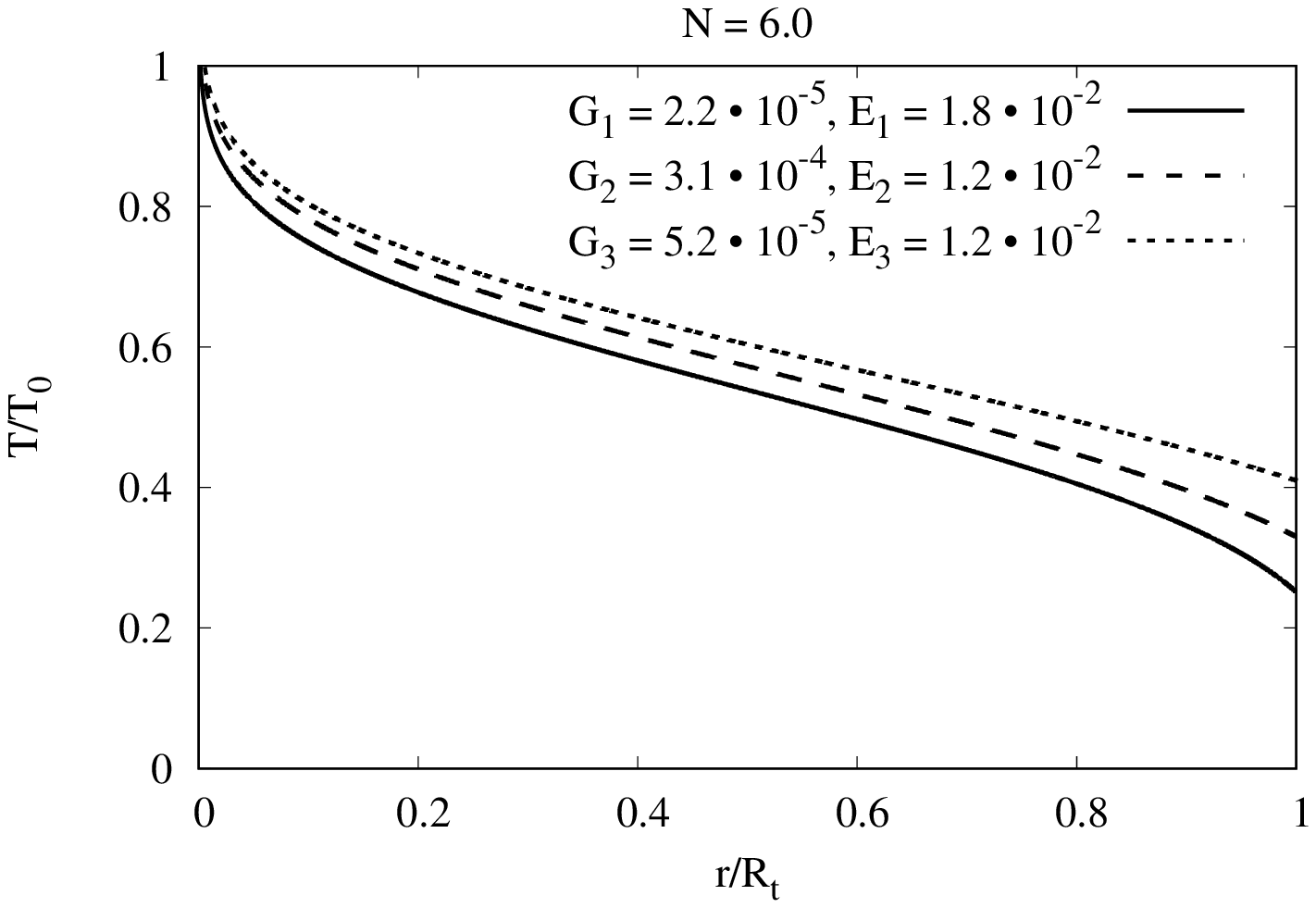}
	\end{center}	
	\caption{Temperature distribution in the  torus small circle for the same parameters as in Fig. 7}\label{fig5a}
\end{figure}

Equations (\ref{syseqforqmodel}) have following form with new parameters:

\begin{equation}
	\frac{db_1}{dx} = N\frac{(1+b_1){\tilde T}^{1/2}}{x}, \qquad
	\frac{d\tilde T}{dx}= -\frac{1+G (1+b_1)^2 {\tilde  T^3}}{x E {\tilde T^{5/2}}}.
	\label{syseqforqmodeln}	
\end{equation}	
We solve equations (\ref{syseqforqmodeln})  numerically in the interval ${\rho_1}/{R_t} = x_1\leq x\leq 1$ at boundary conditions:

\begin{equation}
	b_1(1) = 0, \qquad \tilde T(x_1) = 1, \qquad x_1 = 10^{-4}.
\end{equation}
Results of the solution are presented on the figures (\ref{fig3}-\ref{fig5a})  for the case of plasma parameters in the laboratory facilities.

{The most popular model of quasars, or AGN, consists of the plasma torus, or thick disk, around the supermassive black hole, see i.g. \cite{dk2010}.  The jets, ejected perpendicular to the disk surface, may carry the elongated electrical current, responsible for their collimation.}

	\section{Discussion}                                     
Equations describing magnetic field dynamics in fully ionized nonuniform plasma are derived rigorously, with account of Hall currents and {thermal diffusion} effects. It the system of equation written here, the {self} and externally produced values of magnetic and {electric} fields ${\bf B}$ and ${\bf E}$, and also {electric} current ${\bf j}$ are distinctly separated. 
In presence of {thermal diffusion} the condition for creation of the seed magnetic field in the {unmagnetized} media is found, that is modeling the action of the mechanism, known as "Biermann battery".  
Application of these equations is done for examples of plasma cylinder and plasma torus. In both cases the externally induced electric current may be formed by the temperature gradient ({thermal diffusion}), or by external electric field (battery or accumulator). In all cases the {self} magnetic field, produced by the Hall currents, has a direction opposite to the externally induced magnetic field.

The derived equations could be used for theoretical modeling of magnetic field behavior in astrophysical objects, like different types of stars, fully ionized galactic gas, etc. Another application may be connected with numerical modeling of laboratory experiments with production and acceleration of magnetized, high energy, non-uniform plasma, some of which are related to laboratory  astrophysics.

\end{document}